\documentclass{article}

\usepackage{spconf}
\usepackage{array}
\usepackage{mdwmath}
\usepackage{mdwtab}
\usepackage{fixltx2e}
\usepackage{url}
\usepackage{color}
\usepackage[]{graphicx}
\usepackage{epsfig}
\usepackage{bbm}
\usepackage{enumerate}
\usepackage{amsfonts}
\usepackage{amsmath}
\usepackage{amssymb}
\usepackage{cite}
\usepackage{algorithm}
\usepackage{algpseudocode}
\usepackage{eqparbox}
\usepackage{comment}
\usepackage{breqn}
\usepackage{dblfloatfix}

\providecommand{\abs}[1]{\lvert#1\rvert}
\providecommand{\norm}[1]{\lVert#1\rVert}

\DeclareMathOperator*{\argmin}{argmin}

\title{Wavelength-resolved Neutron Tomography for Crystalline Materials}
\name{S.~V.~Venkatakrishnan$^{\star}$
  \quad Luc~Dessieux $^{\star \dagger}$
  \qquad Philip~Bingham$^{\star}$
  \address{$^{\star}$ Imaging, Signals and Machine Learning Group, Oak Ridge National Lab, Oak Ridge, TN 37831 \\
$^{\dagger}$ Department of Physics, University of Tennessee, Knoxville, TN 37931}
      \thanks{{This manuscript has been authored by UT-Battelle, LLC, under Contract No. DE-AC05-00OR22725 with the U.S. Department of Energy. The United States Government and the publisher, by accepting the article for publication, acknowledges that the United States Government retains a non-exclusive, paid-up, irrevocable, world-wide license to publish or reproduce the published form of this manuscript, or allow others to do so, for United States Government purposes. DOE will provide public access to these results of federally sponsored research in accordance with the DOE Public Access Plan (http://energy.gov/downloads/doe-public-access-plan).
       S.V. Venkatakrishnan was supported by Oak Ridge National Lab via the Laboratory Directed Research and Development program.}}
}

\begin{document}
\maketitle
\begin{abstract}
  Wavelength-resolved (WR) neutron transmission tomography is an emerging technique to characterize engineering materials.
  While tomographic reconstruction for amorphous samples is straightforward, it is challenging to reconstruct samples with single-crystal domains because the attenuation of the sample varies as a function of its orientation with respect to the incident beam due to Bragg scattering.
  In this paper, we present an algorithm that can reconstruct samples with single-crystal domains from WR neutron tomographic measurements. 
  In particular, we use a model-based iterative reconstruction (MBIR) technique that reconstructs the volume by identifying and leaving out the regions of the measurement that are affected by Bragg scatter. 
  We combine the output of the MBIR method with an algorithm that matches the reconstruction to the identified Bragg scatter to reconstruct a feature that corresponds to the local crystallography of the sample being measured.
  Using simulated data, we demonstrate how
  our algorithm can reconstruct materials with single-crystal domains,
  thereby adding a powerful new capability for WR neutron imaging instruments.
\end{abstract}
\section{Introduction}
Neutron imaging is a powerful tool for characterizing engineering materials because they can penetrate thick specimens and provide a complimentary contrast to X-rays~\cite{anderson2009neutron,kardjilov2011neutron}.
Driven by the recent advent of spallation (pulsed) neutron sources and the availability of time-of-flight imaging detectors \cite{tremsin2013high}, there has been an emergence of wavelength-resolved (WR)/time-of-flight (ToF) neutron imaging instruments~\cite{bilheux2015overview,shinohara2016final,kockelmann2018time,nelson2018neutron}.
While WR imaging instruments have been used for tomographic characterization of amorphous samples (for example~\cite{woracek2015neutron}), using them to image samples with single-crystal regions is not common because of Bragg-scatter (that manifests as sharp drops in transmitted intensity) which makes it difficult to invert the measurements.

Engineering materials containing single-crystal domains can be characterized using neutron diffraction~\cite{woracek2018diffraction}. 
However, neutron diffraction involves point-wise raster scanning of a beam across the sample making it extremely time-consuming for large samples. 
This has led to the development of neutron diffraction contrast tomography~\cite{peetermans2014cold} using a pair of detectors, one in a transmission geometry and the other in reflection mode, to reconstruct the shape and crystallographic orientation of individual domains (grains) in a sample. 
Recently, Cereser et al.~\cite{cereser2017time} used a WR neutron imaging detector in transmission geometry to tomographically reconstruct the morphology of large domains (grains) by using an algorithm to detect Bragg scatter events in the measurements and back-projecting them to obtain an approximate shape. 
From a tomographic reconstruction perspective, the central challenge in WR neutron transmission tomography for samples with single-crystal domains is that the these regions diffract the incident beam resulting in measurements that cannot be accurately inverted using analytic algorithms such as the filtered-back projection (FBP).

In this paper, we propose an algorithm for reconstructing samples with single-crystal domains using WR neutron tomography measurements.
In particular, we extend the robust model-based iterative reconstruction
(R-MBIR) of~\cite{VenkatBF15} that casts the reconstruction as minimizing a regularized cost function to the hyper-spectral tomography case. 
In addition to the 3D hyper-spectral volume of the attenuation
coefficients, our algorithm also generates a map of
regions in the original measurements that were identified as anomalous which we refer to as ``Bragg-maps''.
We then develop an algorithm to segment the reconstructed volume and the Bragg-maps and associate each component of the Bragg-map with a segmented domain in the reconstruction.
In contrast to the technique in \cite{VenkatBF15}, our hyper-spectral algorithm enables the reconstruction of a more useful crystallographic signature for each domain in the sample.
In summary, our method makes use of all the measurement data
and can quantitatively reconstruct the sample including
the morphology of the single-crystal domains and a crystallographic signature for each of those domains, thereby adding a vital new capability to WR neutron imaging instruments.

\section{Overall Approach for Wavelength-Resolved Tomography of Crystalline Samples}
\begin{figure}[!htbp]
\begin{tabular}{cc}
    \includegraphics[scale=0.33,trim=0cm 0.15cm 1cm 0cm,clip]{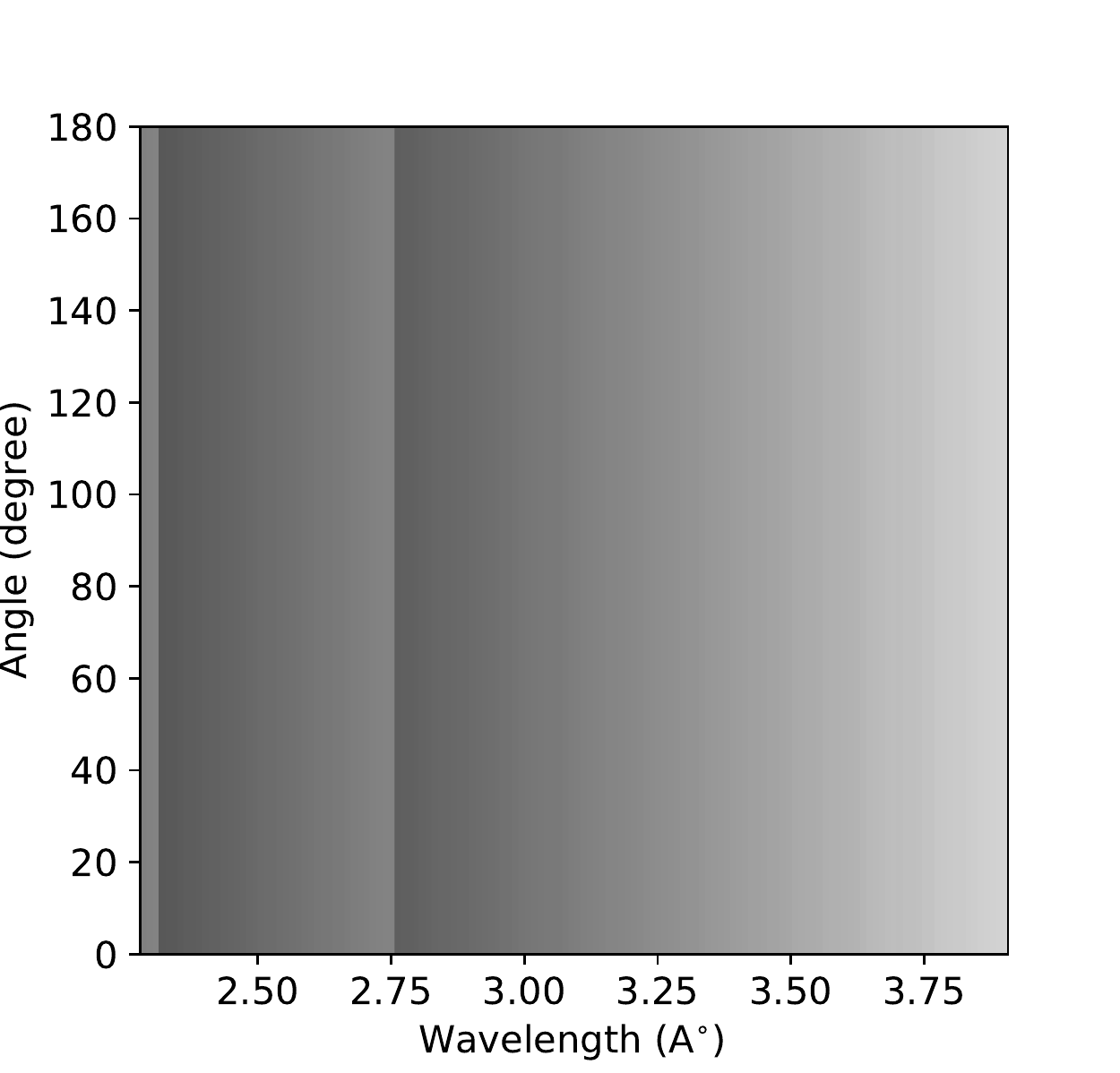}  &
    \includegraphics[scale=0.33,trim=0cm 0.15cm 1cm 0cm,clip]{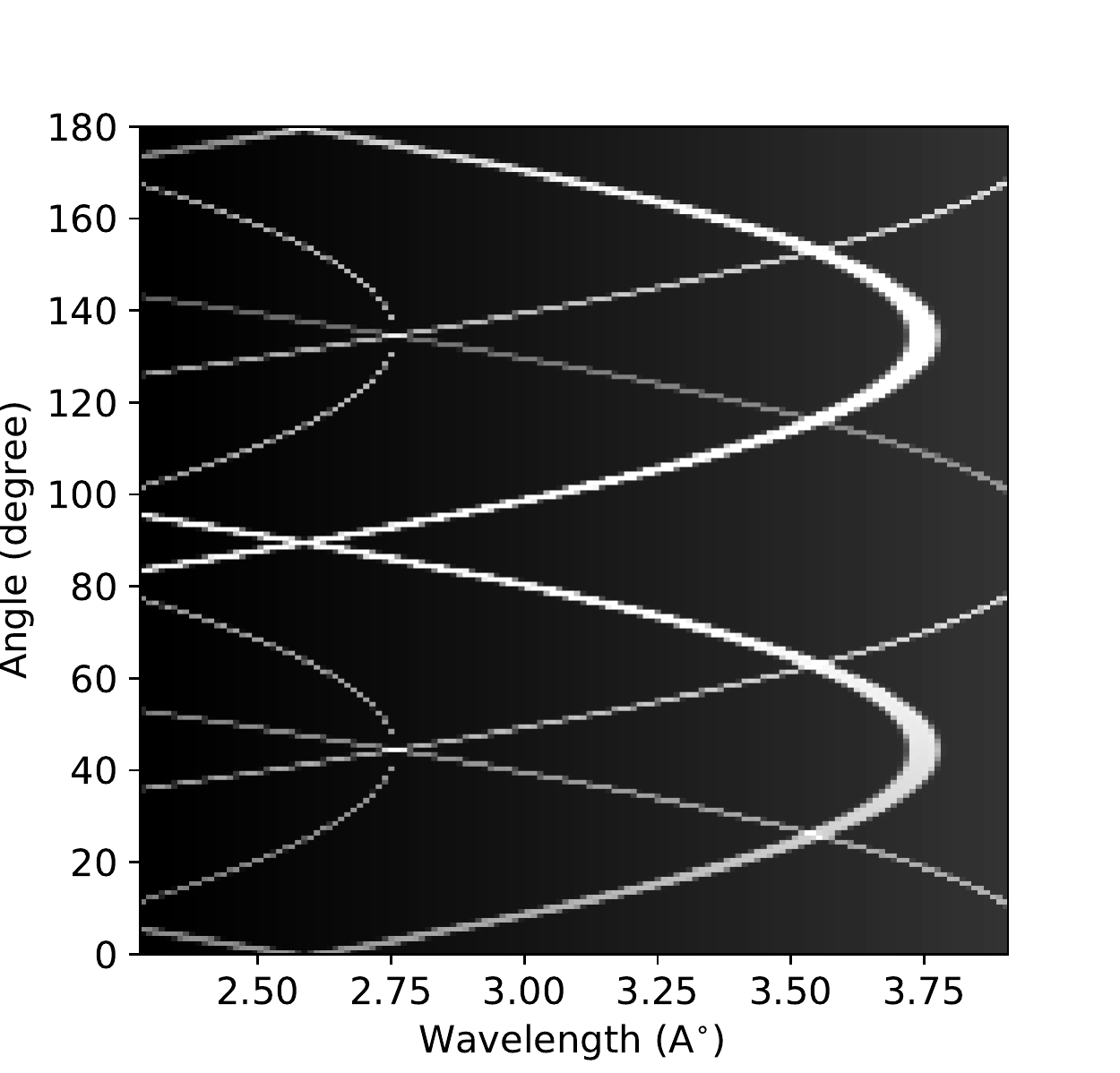} \\
    (a) & (b) 
\end{tabular}
\caption{\label{fig:sinpol_sim} Linear attenuation coefficient of a sample of (a) powder  and (b) single-crystal as a function of rotation angle and wavelength (displayed in range [$5.5,15$] $\times 10^{-5}\mu$m$^{-1}$). For the powder, the attenuation does not change as a function of rotation angle, while for the crystal it changes significantly.}
\end{figure}
In parallel-beam WR neutron tomography, a sample is typically
rotated about a single axis and at each angle a
WR/hyper-spectral projection image is measured.
The measurements corresponding to each wavelength 
are then used to reconstruct a 3D volume
using a tomographic reconstruction algorithm like FBP~\cite{woracek2015neutron}
and then combined to obtain a hyper-spectral tomographic reconstruction.
Extending this approach to characterize samples with single-crystal domains
has been avoided because these regions
diffract the beam when they satisfy the Bragg condition, 
\[ n \lambda = 2d\sin(\theta) \]
where $n$ is some integer, $\lambda$ is the wavelength of
incident neutrons, $d$ is the crystallographic spacing and
$\theta$ is the angle between incident beam and the normal to
the crystallographic plane. 
Specifically, the linear attenuation
coefficient of the single-crystal region changes with respect to its orientation 
to the incident beam for a specific wavelength.
Fig.~\ref{fig:sinpol_sim} shows the simulated attenuation coefficient of a $1$ mm $\times$ $1$ mm $\times 1$ mm \textit{voxel} as a function of wavelength and rotation angle about a single axis generated using \textit{SinPol}~\cite{dessieux2018single},
a WR neutron transmission simulator. 
When the sample is amorphous (also called a \textit{powder}),
the linear attenuation coefficient does not vary as a function of rotation angle as is typical in standard tomography problems. 
However, if the sample is a single-crystal, 
the attenuation of the voxel can change as a function of
the rotation angle because at certain orientations
it satisfies the Bragg condition leading to a sharp increase in attenuation (Fig.~\ref{fig:sinpol_sim}(b)). 
Furthermore, the specific sinusoidal pattern in Fig.~\ref{fig:sinpol_sim}(b) is related to the orientation of the single-crystal with respect to the incident beam.  
 \begin{figure}[!htbp]
 \begin{center}
    \includegraphics[scale=0.32,trim=0cm 0.15cm 2cm 1.25cm,clip]{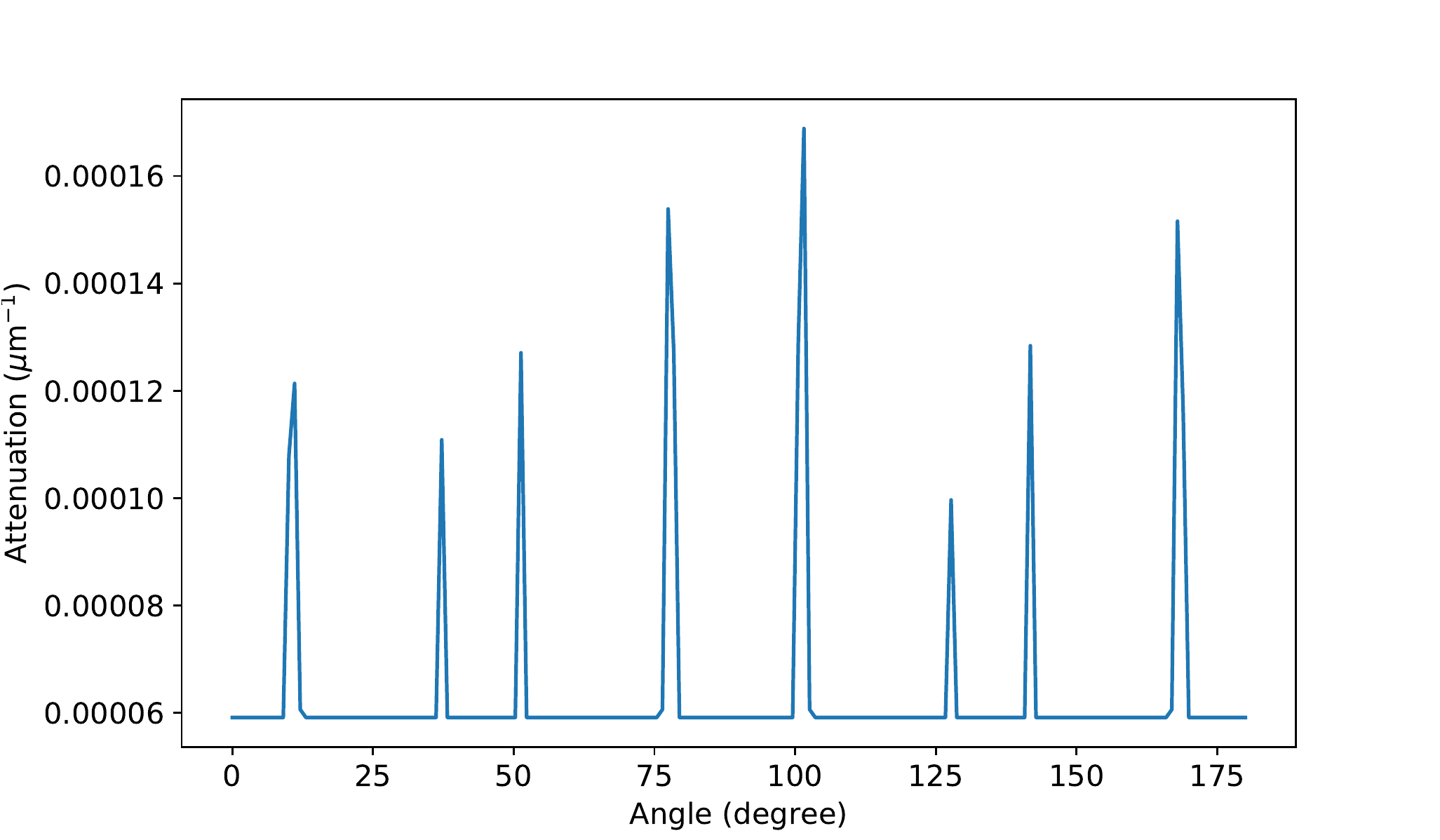}
 \end{center}
 \vspace{-0.2in}
 \caption{\label{fig:crystal_vs_angle} Attenuation coefficient of a single-crystal at $3.25$ A$^{\circ}$ from Fig.~\ref{fig:sinpol_sim}(b) as a function of rotation angle. The attenuation rapidly increases only at a sparse subset of rotation angles. This observation is central to the reconstruction algorithm.}
\end{figure}

\begin{figure*}
\begin{center}
    \includegraphics[scale=0.4,trim=0.5cm 7.5cm 2cm 1.9cm,clip]{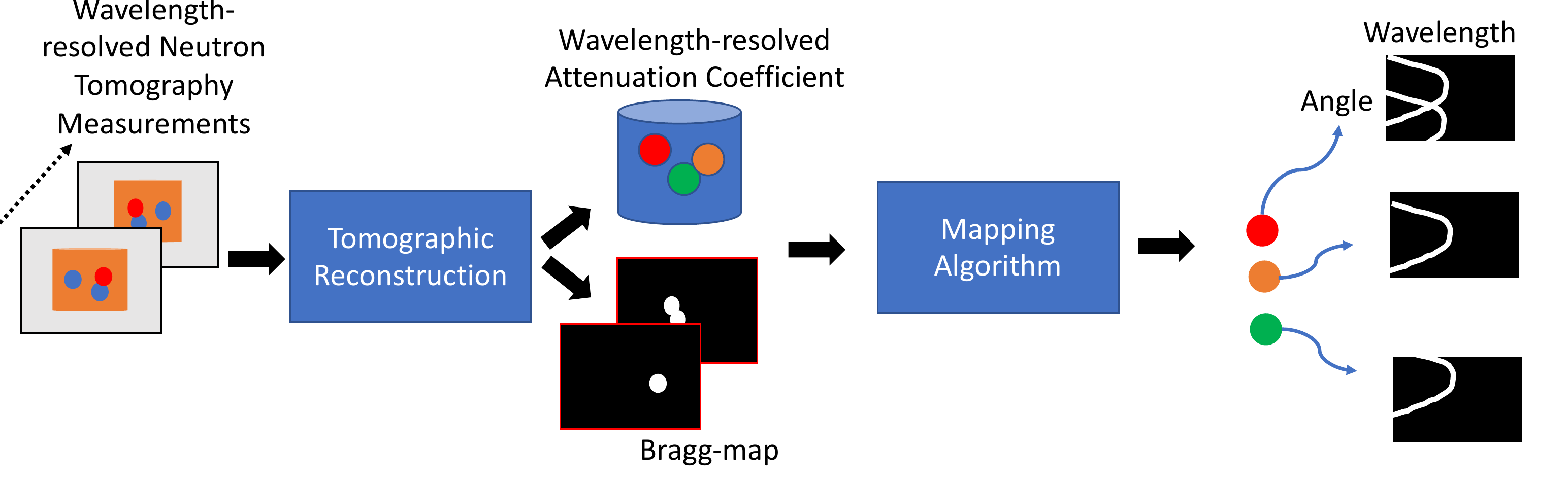}
\end{center}
\vspace{-0.2in}
\caption{\label{fig:overall_algo} Block diagram of the overall algorithm. The input are the WR/hyper-spectral images taken at different rotation angles. The output is the 3D reconstruction of attenuation coefficients, morphology and a binary signature which indicates the local orientation of the crystals in different domains of the 3D reconstruction.}
\end{figure*}

Our goal is to design an algorithm to reconstruct samples with single-crystal domains
in-order to be able to infer the attenuation coefficient,
morphology and some local crystallographic information
from WR tomography measurements. 
While the change of attenuation coefficient with respect to rotation angle would make the reconstruction
problem severely ill-posed in the general case,
the specific pattern in which the attenuation changes
can be exploited to design a reconstruction algorithm.
Intuitively, since the attenuation increases only at a sparse subset of angles
for a given wavelength (see Fig.~\ref{fig:crystal_vs_angle}) we can design a
method to automatically identify those
measurements and leave them out as a part of the tomographic reconstruction. 
These measurements that have been left out constitute the Bragg-map
while the reconstruction gives us the WR attenuation coefficient
and morphology of the sample.
We note that such a reconstruction corresponds to the attenuation coefficient
when the sample is not in the Bragg condition (For example, a value of
$6\times 10^{-5}$ $\mu$m$^{-1}$ in Fig.~\ref{fig:crystal_vs_angle}). 
Finally, we can use the reconstruction and the Bragg-maps to extract a local crystallographic signature for individual domains in the reconstructed volume.  
Our overall approach is summarized in Fig.~\ref{fig:overall_algo}.
In the next two sections, we will outline details of the two blocks that form the core of our proposed approach.

\section{Robust Hyper-spectral Model-based Tomography}
\label{sec:mbir}
In this section, we present a method for wavelength resolved
tomography when the measurements include Bragg scatter 
(observed as sharp drops in the transmitted intensity).
Our approach will simultaneously identify the Bragg-scatter
measurements and reconstruct the hyper-spectral attenuation coefficients in 3D. 
We use the robust model-based iterative
reconstruction (R-MBIR) method of~\cite{VenkatBF15} which has a Huber-like penalty to account for outliers (like Bragg scatter) and
extend it to the hyper-spectral case.
For each wavelength channel $k$, the reconstruction is obtained as
\[
\left(\hat{f}_k,\hat{B}_k\right) \leftarrow \text{R-MBIR}\left(g_{k};T_{k},\sigma_{f_k}\right), 
\]
where $g_k$ is a vector of normalized transmission measurements at wavelength index $k$,
$\hat{f}_k$ is a vector of linear attenuation coefficients of all the voxels in 3D,
$\hat{B}_k$ is a binary Bragg-map of the same dimensions as $g_k$  that indicates
which of the original measurements are anomalous, $T_{k}$ and $\sigma_{f,k}$ are
parameters of the reconstruction algorithm. 
This method has the advantage that it can jointly
identify the anomalous measurements (such as Bragg scatter) and obtain a 3D tomographic reconstruction.

In particular, the reconstruction for a given wavelength using the R-MBIR approach is obtained as 
\begin{eqnarray*}
\hat{f} \leftarrow \argmin_{f} c(f)
\end{eqnarray*}
where 
\begin{dmath}
\label{eq:OriginalCost}
c(f)=\frac{1}{2}  \sum_{i=1}^{M} \beta_{T} \left( (g_{i} - 
[Af]_i)\sqrt{W_{ii}} \right) + R(f;\sigma_{f}) 
\end{dmath}
$A$ is the tomographic projection matrix, $M$ is the total number of measurements,
$W$ is a diagonal $M \times M$ matrix
of inverse noise variance estimated from transmission measurements~\cite{SaBo93},
$R$ is a differentiable edge-preserving 3D regularizer~\cite{VenkatBF15}
with a smoothness parameter $\sigma_{f}$, 
\begin{eqnarray*}
\label{eq:Talwar}
\beta_{T}(x) = \begin{cases}
x^{2} & \abs{x} < T\\
T^{2} & \abs{x} \geq T  
\end{cases}
\end{eqnarray*}
is a generalization of the Huber function~\cite{HuberRobust03,Talwar75}
and $T$ is the outlier threshold.
Using a majorization-minimization approach with quadratic surrogates
along with an optimized gradient method~\cite{VenkatLam17}, we obtain the $3D$ reconstruction
and a map of measurements that have been identified 
as being anomalous (whose normalized fitting error is greater than $T$).

\section{Reconstruction of Crystal Signatures}
\label{sec:analysis}
Given the wavelength-resolved 3D reconstruction, $\left(\hat{f}_{1},...,\hat{f}_{K}\right)$, 
and the Bragg-maps, $\left(\hat{B}_{1},...,\hat{B}_{K}\right)$, 
we use a correlation based approach to associate each of the connected components
in the Bragg-maps to the connected-components in the 3D volume.
This approach will output a binary feature map for each domain (connected component) in the tomographic reconstruction that indicates if it met the Bragg condition at a specific view angle and wavelength, thereby providing information about the local crystallography of the sample.
In essence, this feature vector is a (binary) thresholded version of the attenuation profile in Fig.~\ref{fig:crystal_vs_angle}(b) for each domain in the 3D volume.

In order to find the connected components in the reconstructed volume,
we first apply the K-means segmentation algorithm to the hyper-spectral
reconstructed volume.
A standard connected components analysis is then applied to the segmented volume to
obtain a collection of ($P$) domains.
Next, we apply a connected components analysis to the binary Bragg-map images.
The result of this processing is a set of $Q$ connected components
each associated with a specific view angle $\phi$ and a wavelength $\lambda$.
Finally, for each of the $P$ domains, we forward project it at the different view angles,
binarize the projection
and compute a correlation score~\cite{VenkatBF15} with each of the relevant connected components $Q$.
If $p$ is a binarized version of the projection image of a given particle
and $q$ is the binary anomaly classifier image with the relevant connected component,
we define the correlation score as 
\begin{eqnarray}
\label{eq:SimScore}
S = 1 - \frac{\norm{p^{t} {\bar{q}}}_{1}+\norm{\bar{p}^{t} q}_{1}}{\norm{p}_{1}+\norm{{q}}_{1}}
\end{eqnarray} 
where $\bar{q}$ and $\bar{p}$ refers to the binary complement
operator, $p^{t}$ is the transpose of $p$, and $\|.\|_{1}$ is the $l_{1}$ norm. 
If the correlation score crosses a threshold, we declare a Bragg event.
Thus upon completion of this process for each of the $Q$ anomalies, we have identified which one of the $P$ domains best matched it.
In conclusion, for each of the $P$ domains we have a view-angle and wavelength at which it was potentially in the Bragg-condition. 

\section{Results}
\label{sec:results}
\begin{figure}[!h]
\begin{center}
  \begin{tabular}{cc}
    \includegraphics[scale=0.21,trim={2.5cm 2.2cm 2.9cm 2.4cm},clip]{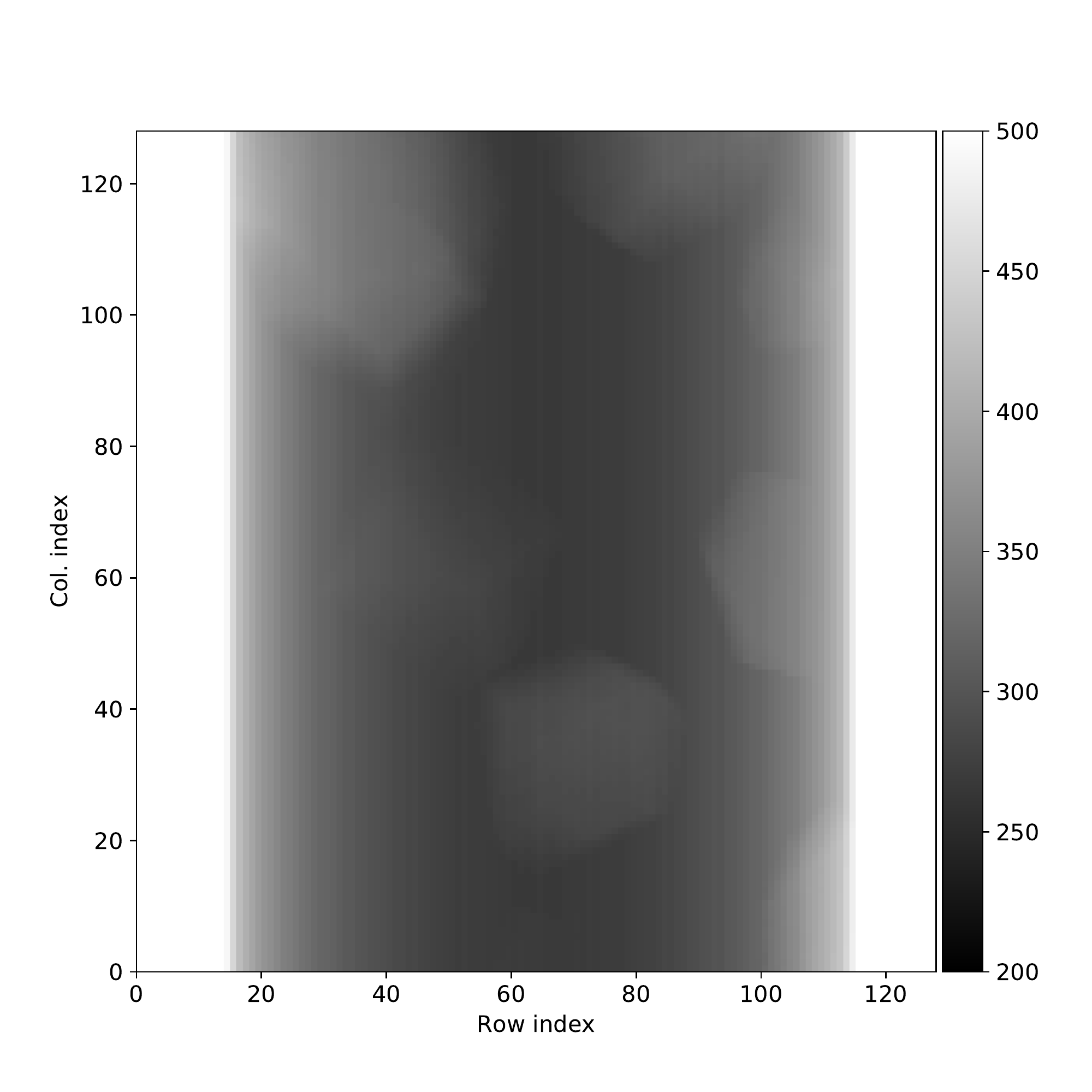} &
    \includegraphics[scale=0.21,trim={2.5cm 2.2cm 2.9cm 2.4cm},clip]{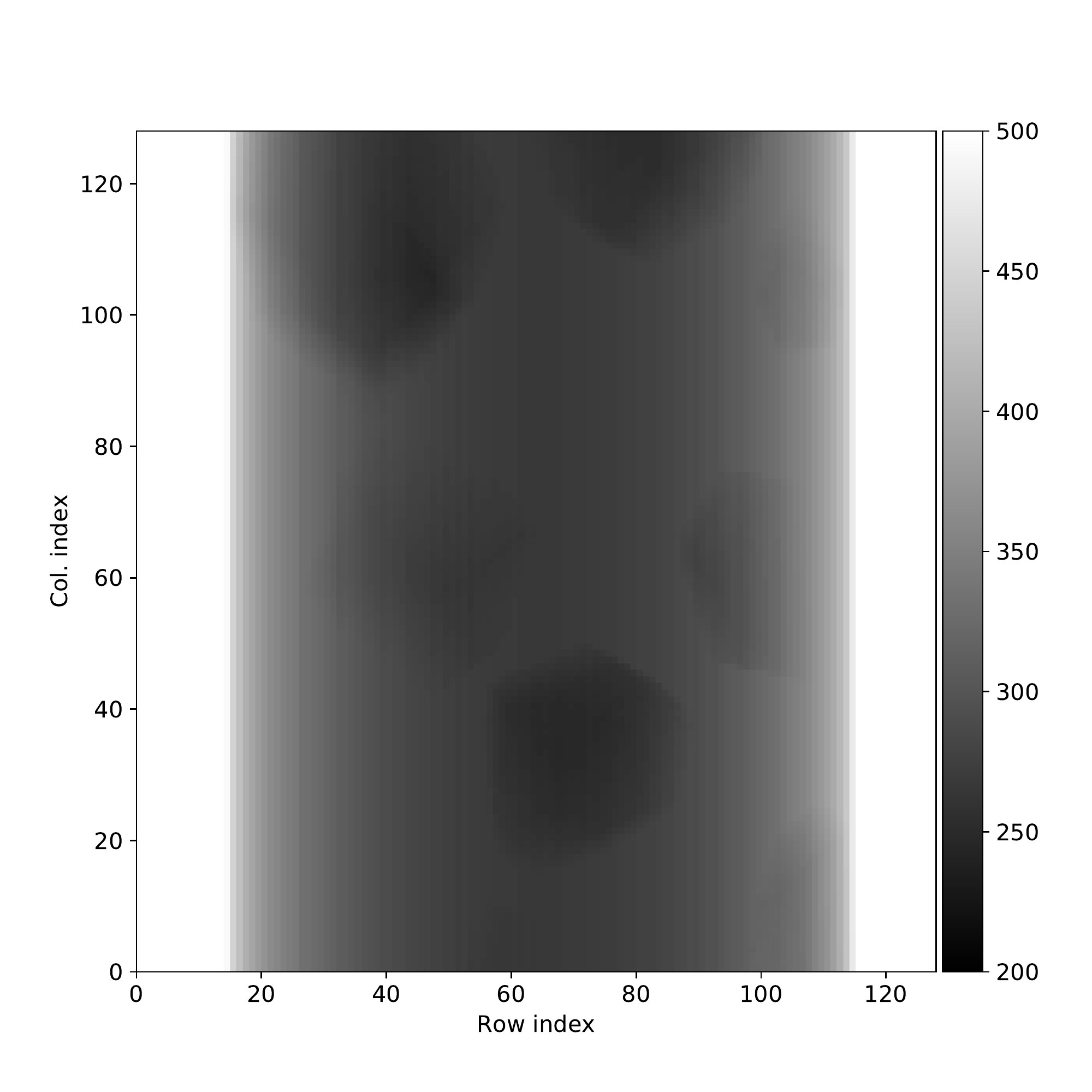} \\
    (a) $\phi$: 35$^{\circ}$, $\lambda$: 3.83 $A^{\circ}$ & (b) $\phi$: 40$^{\circ}$, $\lambda$: 3.83 $A^{\circ}$
\end{tabular}
\end{center}
\vspace{-0.1in}
\caption{\label{fig:simulated_data} Simulated measurements of a cylindrical sample with embedded single-crystal grains for two proximate view angles for a fixed wavelength (displayed in a range [$0,500$] counts). Notice that some of the regions from (a) have a drop in intensity in (b) due to a sharp increase in attenuation of certain regions of the sample due to Bragg scatter.}
\end{figure}
\begin{figure}[!t]
    \includegraphics[scale=0.28,trim=1.8cm 1.0cm 2cm 1.4cm,clip]{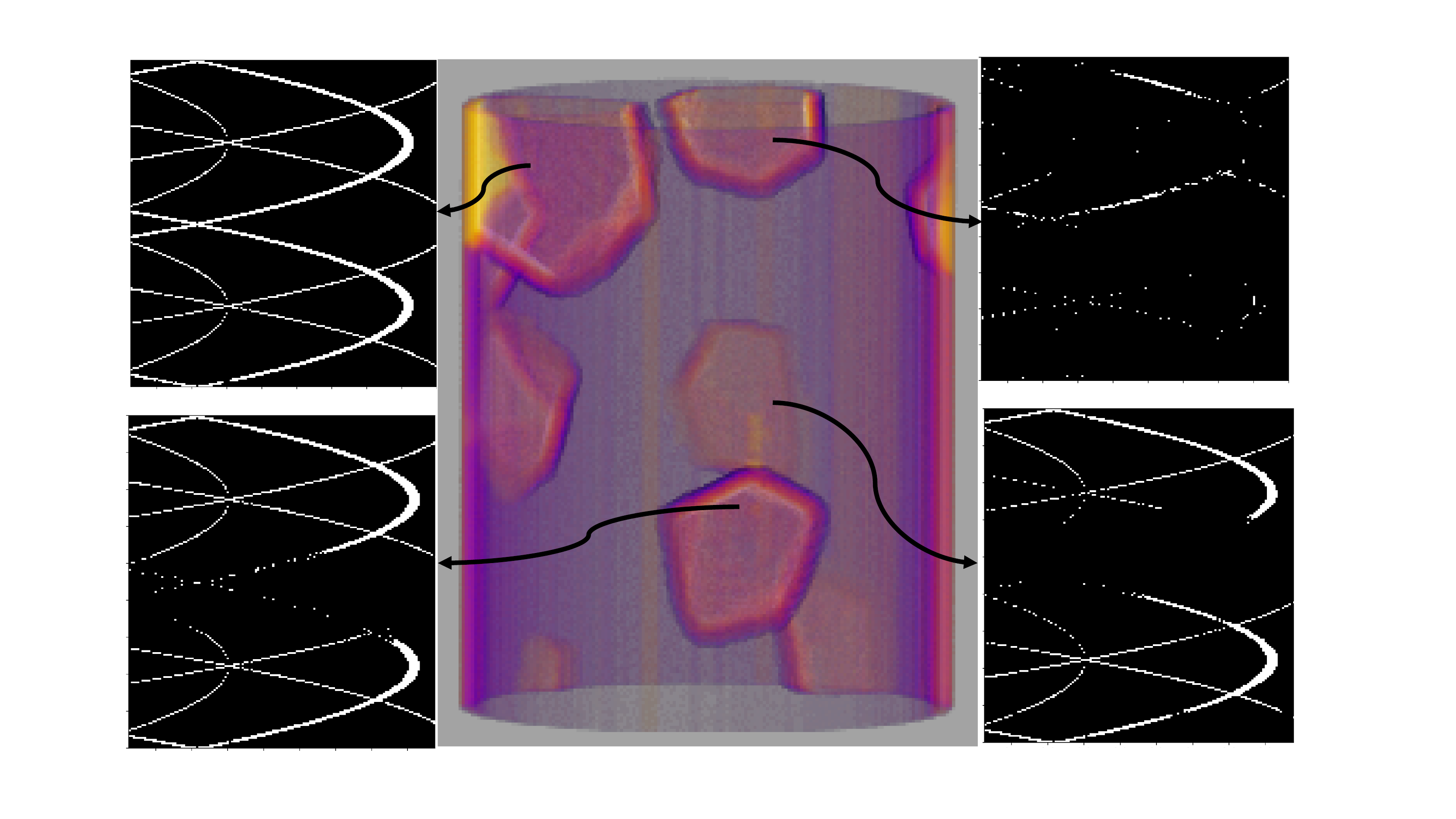}
\vspace{-0.2in}
\caption{\label{fig:recon_crystal_sign} Rendering of the segmented 3D hyper-spectral reconstruction along with
  the reconstruction of the crystallographic signature of the four largest grains. Notice that
  for the largest grains, this signature closely matches the (binarized) ground-truth in Fig.~\ref{fig:sinpol_sim}(b), while for the other grains there are some missed detection due to occlusions.}
\end{figure}

We generate a cylindrical phantom made of a 
powder (attenuation of Fig.~\ref{fig:sinpol_sim} (a)) with embedded single-crystal grains of copper (attenuation coefficient in Fig.~\ref{fig:sinpol_sim} (b))
using the DREAM.3D~\cite{groeber2014dream} software.
All the grains have the same crystallographic orientation with respect to the incident beam. 
We simulate hyper-spectral images of size $128 \times 128$
at $140$ wavelengths between $2.25$ A$^{\circ}$ and $4$ A$^{\circ}$
for $180$ views between $0^{\circ}$ to $180^{\circ}$ using a Beer's law model
applied to the phantom with an incident flux of $I_{0}=500$.
Fig.~\ref{fig:simulated_data} shows the simulated data at two
views for a fixed wavelength.
Notice that some of the regions in Fig.~\ref{fig:simulated_data}(a) are not
distinguishable from the background powder in Fig.~\ref{fig:simulated_data}(b)
because the grains in the sample met the Bragg condition causing a sharp drop in the transmitted intensity.
\begin{figure}[!h]
  \includegraphics[scale=0.37,trim={4.7cm 0cm 3cm 0cm},clip]{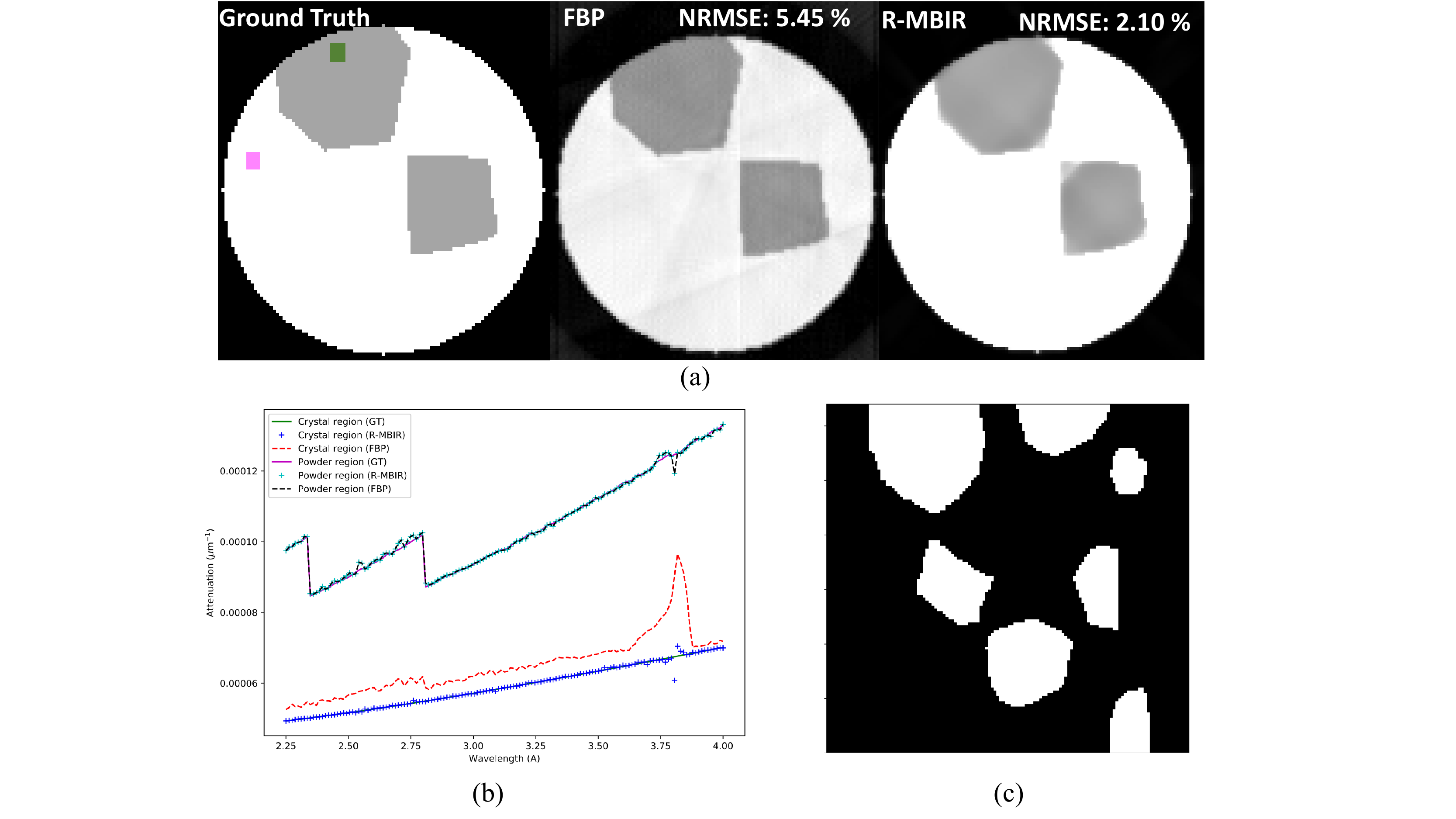} 
\caption{\label{fig:recon_crosssec} Cross section from  (a) ground truth (GT), FBP and R-MBIR method
  at a fixed wavelength ($\lambda=3.83$A) with normalized root mean squared error (NRMSE). (b) Line profiles along wavelength dimension from a powder (cyan square) and a single-crystal region (green square) indicated in the GT image.
  (c) The Bragg-map at a specific view angle and wavelength corresponding to Fig.~\ref{fig:simulated_data}(b), illustrating how the method can correctly identify measurements that were anomalous.}  
\end{figure} 
We reconstruct the sample using cubic voxels with a size of $50\mu$m using the filtered back-projection (FBP) algorithm and the
proposed R-MBIR method. 
For R-MBIR, the value of $T$ is empirically adjusted for each wavelength (based on a rough estimate of the fraction of measurements affected by Bragg scatter at that wavelength), 
while $\sigma_{f}$ is set to $1$ for the q-GGMRF regularizer of~\cite{VenkatLam17}
resulting in a high quality reconstruction with minimal artifacts. 
The threshold for the correlation score is set to $0.5$.
Fig.~\ref{fig:recon_crosssec} shows the output of the tomographic
reconstruction.
Notice that we get a high fidelity reconstruction from R-MBIR
in-spite of the grains strongly diffracting the beam away from the transmission path
at certain orientations, compared to FBP which has strong streaking artifacts. 
Furthermore the R-MBIR method reconstructs a Bragg-map
(for example - Fig.~\ref{fig:recon_crosssec}(c) corresponding to the data in Fig.~\ref{fig:simulated_data}(b)) identifying all the anomalous measurements due to Bragg-scatter, which
conventional methods like FBP do not produce. 
Finally, Fig.~\ref{fig:recon_crystal_sign} shows a 3D rendering of the segmented R-MBIR reconstruction and the reconstructed crystallographic signature of the four largest grains.
While this matches the ground truth (Fig.~\ref{fig:sinpol_sim}(b)) for the large grains (true-positive rates for the $4$ grains in anti-clockwise direction $0.94,0.70,0.63,0.14$), there are missed detections especially for the smaller grains due to the chosen threshold value for the correlation score and challenges in segmenting the Bragg-maps in the case of occlusions.
However, we note that a fitting routine is used to determine the crystallographic
orientation from such signatures~\cite{cereser2017time}, hence a few missed detections will typically not affect the final solution as long as we can determine the parameters from the reconstructed patterns.
\vspace{-0.1in}
\section{Conclusion}
\label{sec:concl}
We presented an algorithm that can quantitatively reconstruct samples with
single-crystal domains along with local crystallographic
orientation information only using WR neutron transmission tomography measurements,  
thereby enabling a new measurement capability.
In the future, we plan to extend our algorithm to cover a broader range of samples such as those that are purely crystalline, and those with the presence of crystallographic texture. 

\bibliographystyle{IEEEbib}
\bibliography{spectralNeutron}

\end{document}